\documentclass[aps,prl,12pt,showpacs,showkeys]{revtex4-1}

\usepackage{graphicx}
\usepackage{epsfig}

\newcommand{\Tr}{{\rm Tr }}
\newcommand{\dl}{\delta}
\newcommand{\ep}{\varepsilon}
\newcommand{\la}{\lambda}
\newcommand{\om}{\omega}

\newtheorem{remark}{Remark}

\begin{document}
\title{Are there traps in quantum control landscapes?}
\author{Alexander N. Pechen\footnote{Present address:
Steklov Mathematical Institute of Russian Academy of Sciences, Gubkina str. 8,
Moscow, 119991, Russia}}
\author{David J. Tannor}
\affiliation{Department of Chemical Physics, Weizmann Institute of Science,
Rehovot 76100, Israel}
\date{}

\begin{abstract}There has been great interest in recent years in quantum control landscapes. Given an objective $J$ that depends on a control field $\ep$ the dynamical landscape is defined by the properties of the Hessian $\dl^2 J/\dl\ep^2$ at the critical points $\dl J/\dl\ep=0$. We show that contrary to recent claims in the literature the dynamical control landscape can exhibit trapping behavior due to the existence of special critical points and illustrate this finding with an example of a 3-level $\Lambda$-system. This observation can have profound implications for both theoretical and experimental quantum control studies.
\end{abstract}
\maketitle

Quantum control aims to manipulate the dynamics of physical processes on the atomic and molecular scale. It is a rapidly growing field of science with numerous applications ranging from selective laser-induced atomic or molecular excitations to high harmonic generation, quantum computing and quantum information, and control of chemical reactions by specially tailored laser pulses, etc.~\cite{qc1,Tannor,qc2,qc3,qc4}.

Generally quantum control problems can be formulated as the maximization of an objective function $J(\ep)$ by a suitable optimal control $\ep$.  A wide variety of quantum control phenomena, selective bond breaking, etc. can be described by control objectives of the form $J(\ep)=\Tr[U_\ep(T)\rho_0 U_\ep^\dagger(T) O]$, where $O$ is an operator describing the target, $\rho_0$ is the initial density matrix and $U_\ep(T)$ is the evolution operator under the action of the control $\ep$ satisfying the equation
\begin{equation}\label{eq6}
 \frac{dU_\ep(t)}{dt}=-i[H_0-\mu\ep(t)]U_\ep(t)\, ,
\end{equation}
where $H_0$ is the free system Hamiltonian and $\mu$ is the dipole moment.

The objective $J=J[\ep]$ as a function of the control $\ep$ defines the landscape of the control problem.  The structure of the landscape determines the complexity of the underlying control problem. Particularly important features of a control landscape are traps --- local maxima of $J(\ep)$.  Traps can have a profound influence on both theoretical and experimental quantum control studies --- they can slow down or even prevent finding globally optimal controls and can lead to erroneous physical conclusions about optimal processes and robustness. We show that contrary to recent claims in the literature~\cite{Science,Landscapes1,Landscapes2,Landscapes3,Landscapes4,Landscapes5} the dynamical control landscape can exhibit trapping behavior due to the existence of special critical points and illustrate this finding with an example of a 3-level $\Lambda$-system. This observation can have profound implications for both theoretical and experimental quantum control studies.

\begin{figure*}[t]
\includegraphics[scale=0.56]{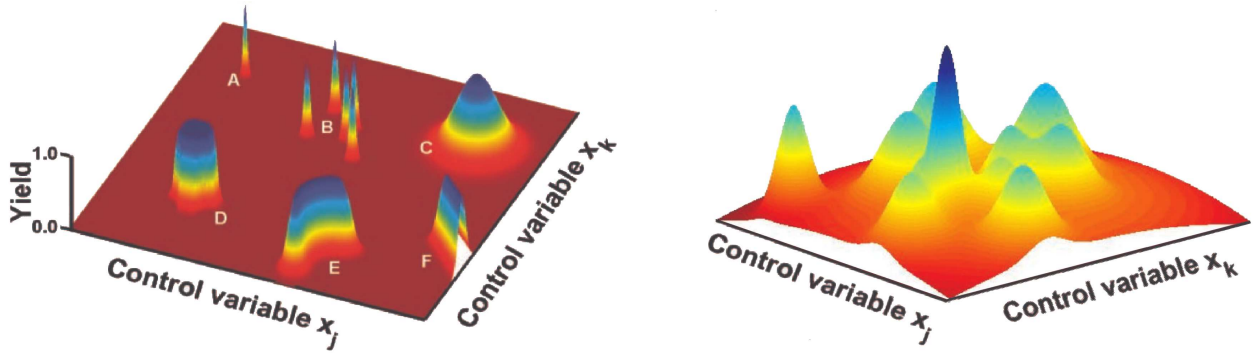}
\caption{\label{fig1} Left: Cartoon of a landscape without traps (local
maxima). All peaks are of the same height and thus all of them are global
maxima. From {\it Science} {\bf 303}, 1998 (2004). Reprinted with permission
from AAAS. Right: Cartoon of a landscape with traps. The landscape has one
highest peak representing the global maximum and several peaks of lower height
representing multiple local maxima. Both landscapes are plotted for two control
variables, $x_j$ and $x_k$, representing the control field $\ep(t)$ at two
different time moments. The actual number of variables in practical applications
may be several hundreds. A local search over the landscape on the left will
eventually reach a global maximum, due to the absence of traps.  However, a
local search over the landscape on the right will most likely find a trap,
ending the search process without ever finding the highest peak.}
\end{figure*}

To understand why traps are significant, consider the generic problem of finding a globally optimal control $\ep_*$ such that $J(\ep_*)=J_{\rm max}=\max\limits_\ep J(\ep)$. Unless the system is extremely simple, numerical or laboratory optimization algorithms generally need to be employed.  The prevailing theoretical methods start from an initial trial control $\ep$ and use gradient and Hessian (first- and second-order) information to explore the neighborhood for a control with better performance.  This new control is then used as a new starting point and the process is iterated.  Experimentally, evolutionary algorithms are commonly used.  While these algorithms are not strictly first- or second-order, each new generation of controls still has a propensity to explore the neighborhood of the previous generation. If the control landscape has traps then first- and second-order algorithms, which effectively are providing only a local search over this landscape, can be prevented from reaching a globally optimal solution $\ep_*$. Thus, the existence or absence of traps is a significant characteristic for any control landscape. Fig.~\ref{fig1} shows landscapes with and without traps.

The analysis of quantum control landscapes was performed in a series of
pioneering
works~\cite{Science,Landscapes1,Landscapes2,Landscapes3,Landscapes4,Landscapes5}
. Extrema of trace functions over unitary and orthogonal groups were also
studied in a different context~\cite{vonNeumann,Brockett} and in the context of
quantum control~\cite{extrema1,extrema2}. The analysis
in~\cite{Science,Landscapes1,Landscapes2,Landscapes3,Landscapes4,Landscapes5}
concluded the absence of traps. In subsequent work it was established that this
conclusion was under the implicit assumption that the Jacobian $\dl U_\ep/\dl
\ep$ has full rank at any point. Although this assumption was shown to be
violated at times~\cite{Wu2009}, is was believed to be generally applicable.
Recently, a particular example of a trap was constructed~\cite{Schirmer2010}.
The present paper significantly advances the field by showing that second order
traps --- points at which the Hessian $H=\dl^2 J/\dl\ep^2$ is negative
semi-definite --- exist in a wide class of quantum control
systems~\cite{second_order}.

We begin our discussion by distinguishing between the dynamic and the kinematic control landscapes.  Until now we have been discussing the functional $J[\ep]$, but one may also consider the simpler functional $J[U]$, where the dependence of $U$ on $\ep$ is suppressed:
\begin{equation}\label{eq:J[U]}
 J_{\rm K}[U]=\Tr[ U\rho_0 U^\dagger O]
\end{equation}
Equation~(\ref{eq:J[U]}) defines the kinematic control landscape. A dynamic critical point (DCP) is defined by $\nabla J(\ep)=\dl J(\ep)/\dl\ep=0$ whereas a kinematic critical point (KCP) is defined by $\nabla J_{\rm K}(U)=\dl J(U)/\dl U=0$, where $\nabla$ denotes gradient. Dynamic and kinematic traps are subopimal maxima for $J[\ep]$ and $J_{\rm K}[U]$, respectively.

Assuming complete controllability, i.e. that $U$ in~(\ref{eq:J[U]}) can be any unitary operator, the {\it kinematic} control landscape is known to be free of traps: all critical points of $J_{\rm K}[U]$ are either global maxima and minima, or saddles~\cite{trapfree}. This result implies that the {\it dynamic} control landscape will be trap-free if one additionally assumes that the Jacobian $\dl U_\ep(T)/\dl\ep$ has full rank  at any $\ep$~\cite{Ho2006,Raj2007}. Indeed, by the chain rule,
\[
 \frac{\dl J(\ep)}{\dl\ep(t)}=\frac{\dl J_{\rm K}[U_\ep(T)]}{\dl
U_\ep(T)}\frac{\dl U_\ep(T)}{\dl\ep(t)}
\]
and hence under the full rank condition all DCP are at KCP and have exactly the same critical point structure as the corresponding KCP~\cite{Wu2008}.

Our first result concerns the inequivalence of critical point structures. To find the condition for a KCP of~(\ref{eq:J[U]}) we take any infinitesimal variation of $U$ in the form $U\to U'=U(1+\dl U)$~\cite{trapfree}. Unitarity of $U'$ up to the first order in $\dl U$ implies $\dl U^\dagger=-\dl U$, i.e. $\dl U$ is anti-Hermitian, and hence the variation of the objective $J_{\rm K}$ with respect to $U$ is $\dl J_{\rm K}= J_{\rm K}[U']-J_{\rm K}[U]= \Tr \{\dl U[\rho_0, O_T]\}+o(\|\dl U\|)$, where $O_T=U^\dagger O U$. If $U$ is a critical point for $J_{\rm K}[U]$, then the condition $\dl J_{\rm K}=0$ needs to hold for any anti-Hermitian $\dl U$, implying that
\begin{equation}\label{eq1:1}
 [\rho_0, O_T]=0\,.
\end{equation}
Equation~(\ref{eq1:1}) is the condition for a KCP. All $U$ satisfying the
condition~(\ref{eq1:1}) were shown to be either global maxima, minima, or
saddles~\cite{trapfree}, and therefore second order traps do not exist for
kinematic control landscapes. Dynamical critical control fields that violate
(\ref{eq1:1}) were shown to exist for the problem of optimal population transfer
between two pure states of a quantum systems~\cite{Schirmer2010}. We have been
able to generalize this finding by showing that such fields exist not only for
optimal population transfer between two pure states, but for maximizing the
expectation value of a more general class of observables. The proof is given in
Section 3 of the Appendix. This inequivalence of the critical points in the
dynamic and kinematic landscapes is an indication of the breakdown of the full
rank assumption.

We now turn to our main result.  For a general class of systems there exist
dynamical critical controls ($[\rho_0, O_T]=0$) that are second order traps due
to the violation of the full rank assumption. In particular, second order traps
appear in the dynamical control landscape whenever the dipole moment satisfies
$\mu_{ij}=0$ for some $i\ne j$ (i.e. if a direct transition between some pair of
levels is forbidden). In this case there exists an initial density matrix and a
target operator such that $\ep(t)=0$ is a second order trap. (Note that the
condition $\mu_{ij}=0$ can be consistent with the assumption of complete
controllability of the system provided that the levels $i$ and $j$ are connected
indirectly through other states.) More generally, a control $\ep(t)=\ep_0$ is a
second order dynamical trap if in terms of the spectral decomposition $\tilde
H_0=H_0-\mu\ep_0=\sum_{i=1}^n \tilde h_i|\tilde i\rangle\langle \tilde i|$ the
initial density matrix and target operator have the form $\rho_0=|\tilde
k\rangle\langle \tilde k|$ and $O= \sum_{i=1}^n \la_i|\tilde i\rangle\langle
\tilde i|$, where $1<k<n$ and $\la_1>\la_2>\dots>\la_n$, and the dipole moment
satisfies $\langle \tilde i|\mu|\tilde k\rangle=0$ for all $i<k$. (Again, the
condition that $\langle \tilde i|\mu|\tilde k\rangle=0$ for any $i<k<n$ can be
consistent with the controllability assumption if the dipole moment connects the
state $|\tilde k\rangle$ with all states $|\tilde i\rangle$ for $i<k$ through
other states.) To prove this finding, we explicitly compute the Hessian and show
that it is negative semidefinite under the above conditions; the details of the
proof are given in Section 2 of the Appendix.

\begin{figure}
\includegraphics[scale=0.7]{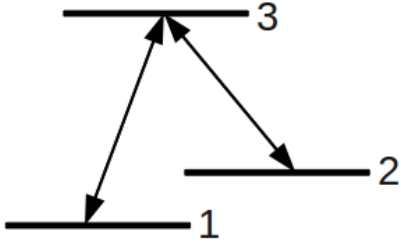}
\caption{\label{scheme}The simplest example of a quantum system possessing a second order trap is a 3-level $\Lambda$-system initially in the ground state. The control field $\ep(t)=0$ is a second order trap for maximizing expectation of any target operator of the form $O=\sum_{i=1}^3\la_i|i\rangle\langle i|$ with $\la_2>\la_1>\la_3$.}
\end{figure}

The simplest example of such a second order trap appears in the problem of
maximizing the expectation of an operator $O=\sum_{i=1}^3\la_i|i\rangle\langle
i|$ with $\la_2>\la_1>\la_3$ for a three-level $\Lambda$-atom initially in the
state $\rho_0=|1\rangle\langle 1|$ (Figure~\ref{scheme}). The dipole moment
for $\Lambda$-atom satisfies $\mu_{12}=0$, consistent with the controllability
assumption if $\mu_{13}\ne 0$ and $\mu_{23}\ne 0$. Globally optimal control
fields steer $|1\rangle$ completely into $|2\rangle$ producing the global
maximum of the objective with value $J_{\rm max}=\la_2$. The control field
$\ep(t)=0$ produces a second order trap with the objective value $J=\la_1<J_{\rm
max}$.

In conclusion, we have established that second order traps in quantum control landscapes exist in a wide range of quantum systems. More research will be required to establish if these points are true traps, but for the local search algorithms currently in use second order traps pose virtually all the same numerical and experimental difficulties as true traps.  Moreover, since the present work establishes that the full rank assumption is violated for a wide class of quantum systems, the previous claims of the absence of traps, which were based on this assumption, have to be completely rethought.

\begin{acknowledgments} This work was supported by the EU Marie Curie Network
EMALI, the Minerva Foundation and the NSF under Grant No. PHY05-51164, KITP
preprint number: NSF-KITP-11-031. This research is made possible by the historic
generosity of the Harold Perlman family.
\end{acknowledgments}

\section{Appendix}
\subsection*{1. Gradient and Hessian for $J(\ep)$} Taking a small
variation of the control $\ep\to\ep+\dl\ep$ gives the following expansion for
the objective $J(\ep)=\Tr[U_\ep(T)\rho_0 U_\ep^\dagger(T) O]$:
\begin{eqnarray*}
 J(\ep+\dl\ep)&=&J(\ep)+\int\limits_0^T dt\dl\ep(t)\frac{\dl J}{\dl\ep(t)}
+\frac{1}{2}\int\limits_0^Tdt_1\int\limits_0^Tdt_2\dl\ep(t_1)\frac{\dl^2
J}{\dl\ep(t_1)\dl\ep(t_2)}\dl\ep(t_2)+o(\dl\ep^2)\\
 &=&\vphantom{\int\limits_0^T}J(\ep)+\langle\nabla J_\ep,\dl\ep\rangle
+\frac{1}{2}\langle\dl\ep,H_\ep \dl\ep\rangle+o(\dl\ep^2)=J(\ep)+\langle\nabla
J_\ep,\dl\ep\rangle +\frac{1}{2}h(\dl\ep)+o(\dl\ep^2)
\end{eqnarray*}
where $\nabla J_\ep$ and $H_\ep$ are the gradient and Hessian of $J$ evaluated
at $\ep$ and  $h(\dl\ep)=\langle\dl\ep,H_\ep \dl\ep\rangle$. Denoting
$V(t)=U_\ep^\dagger(t)VU_\ep(t)$ and $O_T=U^\dagger_\ep(T)OU_\ep(T)$, we may
obtain expressions for the gradient and Hessian of $J$:
\begin{eqnarray}
 \nabla J_\ep(t)&=& \frac{\dl J}{\dl\ep(t)}=-i\Tr\left\{ [\rho_0,O_T]
V(t)\right\}\label{gradient}\\
 H_\ep(t_1,t_2)&=&\frac{\dl^2 J}{\dl\ep(t_1)\dl\ep(t_2)}=\Tr\Bigl\{O_T\Bigl[ 2
V(t_1)\rho_0 V(t_2) -\hat T[V(t_1)V(t_2)]\rho_0-\rho_0\hat T_{\rm
a}[V(t_1)V(t_2)]\Bigr]\Bigr\}\, ,\qquad\label{hessian}
\end{eqnarray}
where $\hat T$ and $\hat T_{\rm a}$ stand for the operators of chronological and
anti-chronological ordering, respectively.

\subsection*{2. Second order dynamical traps at KCP} A DCP $\ep$ satisfies
$\nabla J_\ep=0$. Eq.~(\ref{gradient}) implies that a control field $\ep$ is a
DCP if and only if for any $t\in[0,T]$:
\begin{equation}\label{eq4}
 \Tr\{[\rho_0,O_T]V(t)\}=0
\end{equation}
If $\ep$ is a kinematically critical control, then $[\rho_0,O_T]=0$ and the
Hessian~(\ref{hessian}) at $\ep$ takes the form
\[
H_\ep(t_1,t_2)=\Tr\Bigl\{\Bigl[2V(t_1)\rho_0
V(t_2)-[V(t_1)V(t_2)+V(t_2)V(t_1)]\rho_0\Bigr]O_T\Bigr\}
\]
Then
\begin{eqnarray*}
h(f)=\Tr[{\cal L}_{V_f}(\rho_0)O_T]
\end{eqnarray*}
where $V_f=\int_0^Tf(t)V(t)dt=\int_0^Tf(t)U^\dagger(t)VU(t)dt$ and ${\cal
L}_{V_f}$ is the Lindblad-type superoperator of the form
\[
{\cal L}_{V_f}(\rho)=2 {V_f}\rho {V_f}- {V_f}^2\rho-\rho {V_f}^2
\]

The condition $[\rho_0, O_T]=0$ implies the existence of an orthonormal basis
$|\phi_k\rangle$ such that
\begin{eqnarray}
 O_T&=&\sum\limits_{k=1}^n \la_k|\phi_k\rangle\langle\phi_k|,\qquad \la_{\rm
max}:=\la_1\ge \la_2\ge\cdots\ge\la_n=:\la_{\rm min}\\
 \rho_0&=&\sum\limits_{k=1}^n\om_k|\phi_k\rangle\langle\phi_k|,\qquad \om_k\ge
0,\qquad\sum\limits_{k=1}^n\om_k=1
\end{eqnarray}
Then at a KCP
\begin{equation}\label{eq:h(f)}
  h(f) =\sum\limits_{k=1}^n\om_k\langle f_k|O_T- \la_k |f_k\rangle
\end{equation}
where $|f_k\rangle=V_f|\phi_k\rangle$. The expression~(\ref{eq:h(f)}) can be
rewritten as:
\begin{eqnarray*}
h(f)=\vphantom{\left|\int\right|^2}\sum\limits_{k,i}
\om_k(\la_i-\la_k)|\langle\phi_k|V_f|\phi_i\rangle|^2
\end{eqnarray*}
and can be represented as the difference of two non-negative quantities,
\[
 h(f)=h^+(f)-h^-(f)\, ,
\]
where
\begin{eqnarray*}
 h^+(f)&=&\sum\limits_{1\le i<k\le
n}\om_k(\la_i-\la_k)|\langle\phi_k|V_f|\phi_i\rangle|^2\ge 0\\
 h^-(f)&=&\sum\limits_{n\ge i>k\ge
1}\om_k(\la_k-\la_i)|\langle\phi_k|V_f|\phi_i\rangle|^2\ge 0
\end{eqnarray*}
If the initial state is $\rho_0=|\phi_1\rangle\langle\phi_1|$, corresponding to
the global maximum of the objective, then $h(f)=-h^-(f)=\langle f_1|O_T-
\la_{\rm max} |f_1\rangle\le 0$ and the Hessian is negative semi-definite.
Similarly, if the initial state is $\rho_0=|\phi_n\rangle\langle\phi_n|$,
corresponding to the global minimum of the objective, then $h(f)=h^+(f)=\langle
f_n|O_T- \la_{\rm min} |f_n\rangle\ge 0$ and the Hessian is positive
semi-definite.

If a control field $\ep$ is a DCP at a KCP and satisfies (i) $J(\ep)<J_{\rm
max}$ and (ii) for all $i,k$ such that $\om_k\ne 0$ and $\la_i>\la_k$, for all
times $t\in(0,T)$
\begin{equation}\label{eq:sot}
 \langle\phi_i|V(t)|\phi_k\rangle=0
\end{equation}
then $\ep$ is a second order dynamical trap. Indeed, in this case for any $f$
\[
 h^+(f)=\sum\limits_{\la_i>\la_k}\om_k(\la_i-\la_k)\left|\int\limits_0^T dt
f(t)\langle\phi_i|V(t)|\phi_k\rangle\right|^2=0
\]
and hence $h(f)\le 0$ for any $f$. A similar analysis shows that if the
condition~(\ref{eq:sot}) is satisfied for $\la_k>\la_i$, then $h^-(f)=0$ and
therefore $h(f)\ge 0$.

This result leads to the main statement of this work that a control
$\ep(t)=\ep_0$ is a second order dynamical trap if in terms of the spectral
decomposition $\tilde H_0=H_0-\mu\ep_0=\sum_{i=1}^n \tilde h_i|\tilde
i\rangle\langle \tilde i|$ the initial density matrix and target operator have
the form $\rho_0=|\tilde k\rangle\langle \tilde k|$ and $O= \sum_{i=1}^n
\la_i|\tilde i\rangle\langle \tilde i|$, where $1<k<n$ and
$\la_1>\la_2>\dots>\la_n$, and the dipole moment satisfies $\langle \tilde
i|\mu|\tilde k\rangle=0$ for all $i<k$. Indeed, in this case $O_T=O$ and
therefore $[\rho_0,O_T]=0$. Hence the control $\ep(t)=\ep_0$ is a KCP. The
dipole moment evolves as
\[
 \mu(t)=e^{it\tilde H_0}\mu e^{-it\tilde H_0}
\]
and therefore for any $i<k$
\[
 \langle \tilde i|\mu(t)|\tilde k\rangle=e^{it(\tilde h_i-\tilde h_k)} \langle
\tilde i|\mu|\tilde k\rangle=0
\]
Hence for any $f$
\[
 h^+(f)= \sum\limits_{i=1}^{k-1}(\la_i-\la_k)|\langle \tilde i|\mu_f|\tilde
k\rangle|^2=0
\]
and the Hessian at $\ep(t)=\ep_0$ is negative semi-definite [$h(f)\le 0$]
showing that $\ep(t)=\ep_0$ is a second order trap. In particular, for a 3-level
$\Lambda$-system evolving under the action of zero control $\ep(t)=0$, $\langle
1|\mu(t)|2\rangle=e^{it(h_1-h_2)} \mu_{12}=0$
and therefore for any $f$, $h^+(f)\equiv (\la_2-\la_1)|\langle
1|\mu_f|2\rangle|^2=0$. Thus, for this system $\ep(t)=0$ is a second order trap.

\subsection*{3. Dynamical critical points not at KCP}
The criterion~(\ref{eq4}) can be rewritten in the following equivalent forms
(where the spectral decomposition
$\rho_0=\sum\om_i |\phi_i\rangle\langle\phi_i|$ for the initial state is used in
the second line)
\begin{eqnarray}
 \Im \Tr[\rho_0 O_TV(t)] &=&0\label{eq5} \\
 \sum\limits_{i=1}^n\om_i\Im
[\langle\phi_i|O_TV(t)|\phi_i\rangle]&=&0\label{eq2}
\end{eqnarray}

At a KCP, $\rho_0$ and $O_T$ commute. This implies that they have a common basis
$|\phi_i\rangle$ such that $O_T|\phi_i\rangle=\la_i|\phi_i\rangle$. Since $V(t)$
is Hermitian, the criterion~(\ref{eq2}) is trivially satisfied, i.e. every KCP
is a DCP. For DCP that are not at KCP, $\rho_0$ and $O_T$ do not commute.
Therefore at least one eigenvector of $O_T$ is not an eigenvector of $\rho_0$
and verification of the criterion~(\ref{eq2}) becomes more complicated.
Nevertheless, as we now show the dynamical landscapes can have a critical point
-- a DCP -- when the kinematical landscape does not have a KCP.

Consider a completely controllable $n$-level quantum system with the Hamiltonian
$H=H_0-\mu\ep(t)$,
where $H_0=\sum_{i=1}^n h_i |i\rangle\langle i|$. Moreover, consider initial
states and target operators of the forms
\begin{eqnarray*}
  \rho_0&=&|\psi\rangle\langle\psi|,\qquad
|\psi\rangle=\frac{|i\rangle+e^{i\psi}|j\rangle}{\sqrt{2}}\\
 O&=&e^{-iTH_0}(|\phi\rangle\langle\phi|+Q)e^{iTH_0},\qquad
|\phi\rangle=\frac{|i\rangle+e^{i\phi}|j\rangle}{\sqrt{2}},
\end{eqnarray*}
where $T$ is the final time, $Q$ is any Hermitian operator whose kernel contains
the vector $|\psi\rangle$, and $\alpha=\phi-\psi\ne 0,\pi$.
If $\mu_{ii}=\mu_{jj}$, then the control $\ep(t)=0$ is a DCP that not a KCP.

To prove this statement, first we show that $\hat Z:=[\rho_0,O_T]\ne 0$. In
fact, for $\ep(t)=0$
\[
 O_T=e^{iTH_0} O e^{-iTH_0}=|\phi\rangle\langle\phi|+Q
\]
and therefore
\[
 \hat
Z=[|\psi\rangle\langle\psi|,|\phi\rangle\langle\phi|]=\frac{1+e^{i\alpha}}{2}
|\psi\rangle\langle\phi|-
 \frac{1+e^{-i\alpha}}{2} |\phi\rangle\langle\psi|
\]
where we have used the fact that $[|\psi\rangle\langle\psi|,Q]=0$ due to the
assumption that vector $|\psi\rangle$ is in the kernel of $Q$. This operator is
non-vanishing; for example
\[
 \langle\psi|\hat Z|\phi\rangle=\frac{1+e^{i\alpha}}{2}\frac{1-\cos\alpha}{2}\ne
0 \mbox{ for } \alpha\ne 0,\pi
\]
Now we will prove that $\ep(t)=0$ is a critical control. The
condition~(\ref{eq5}) takes the form
\begin{equation}\label{eq3}
 \Im[\langle\psi|\phi\rangle\langle\phi|\mu(t)|\psi\rangle]=0\qquad\forall
t\in[0,T]
\end{equation}
One has
\[
 \Im[\langle\psi|\phi\rangle\langle\phi|V(t)|\psi\rangle]=\frac{1}{4}(\mu_{ii}
-\mu_{jj})\sin\alpha
\]
Since for our system $\mu_{ii}=\mu_{jj}$, the condition~(\ref{eq3}) is satisfied
and therefore $\ep(t)=0$ is a non-kinematically critical control. The objective
value produced by the control $\ep(t)=0$ is $J(\ep=0)=\Tr[\rho_0
O_T]=|\langle\psi|\phi\rangle|^2=(1+\cos\alpha)/2$,
$J_{\rm min}\le 0<J(\ep=0)<1\le J_{\rm max}$, where $J_{\rm min}$ and $J_{\rm
max}$ are the minimal and maximal values of the objective. Therefore, the
control $\ep(t)=0$ is not a global maximum or minimum, but a DCP that is not a
KCP.

\begin{remark} The case with $Q=0$ corresponding to the problem of optimal
population transfer between two pure states was considered
in~\cite{Schirmer2010}.\end{remark}

\begin{remark} The condition $\mu_{ii}=\mu_{jj}$ for some $i,j$ can be
consistent with the controllability assumption. Moreover, for many physical
systems the diagonal matrix elements of the dipole moment vanish such that this
condition is trivially satisfied. Therefore, this theorem describes a class of
completely controllable systems which under physically reasonable assumptions
about the dipole moment have a local DCP that are not at a KCP.
\end{remark}

\begin{remark} The theorem can be generalized to constant non-zero controls
$\ep(t)=\ep_0$. Let $\tilde H_0:=H_0-\mu\ep_0=
\sum\limits_{i=1}^n \tilde h_i |\tilde i\rangle\langle\tilde i|$. If for some
$i\ne j$, $\langle\tilde i|\mu|\tilde i\rangle=
\langle\tilde j|\mu|\tilde j\rangle$, then the control $\ep(t)=\ep_0$ is
non-kinematically critical for any initial state
and target operator of the forms
\begin{eqnarray*}
  \rho_0&=&|\psi\rangle\langle\psi|,\qquad |\psi\rangle=\frac{|\tilde
i\rangle+e^{i\psi}|\tilde j\rangle}{\sqrt{2}}\\
 O&=&e^{-iT\tilde H_0}(|\phi\rangle\langle\phi|+Q)e^{iT\tilde H_0},\qquad
|\phi\rangle=\frac{|\tilde i\rangle+e^{i\phi}|\tilde j\rangle}{\sqrt{2}},
\end{eqnarray*}
where $T$ is the final time, $Q$ is any Hermitian operator whose kernel contains
vector $|\psi\rangle$, and
$\alpha=\phi-\psi\ne 0,\pi$.
\end{remark}


\begin{thebibliography}{99}
\bibitem{second_order} Second order traps, where the Hessian $H_\ep$ is negative
semi-definite, are not necessarily suboptimal maxima: if the kernel of $H_\ep$
is non-empty and the third or higher order variations of $J$ are non-zero along
a $\dl\ep\in{\rm ker}\,H_\ep$, then $\ep$ may be neither a local maximum nor
minimum. They are however important as being traps for commonly used algorithms
exploiting only the first and second order information.
\end{thebibliography}
\end{document}